\documentclass[aps,prd,superscriptaddress,showpacs,nofootinbib,notitlepage,twocolumn]{revtex4-1}
\usepackage{graphicx,subfigure,amsmath,amsfonts,amssymb,multirow,mathrsfs}
\usepackage[colorlinks,linkcolor=blue,citecolor=blue,urlcolor=blue]{hyperref}

\def\aap{Astron.\ Astrophys.\ }
\def\apj{Astrophys.\ J.\ }
\def\apjl{Astrophys.\ J.\ Lett.\ }

\def\mnras{Mon.\ Not.\ Roy.\ Astron.\ Soc.\ }
\def\physrep{Phys.\ Rept.\ }
\def\prd{Phys.\ Rev.\ D\ }
\def\prl{Phys.\ Rev.\ Lett.\ }

\def\nphysa{Nucl.\ Phys.\ A\ }

\begin{document}
\newcommand{\PMO}{Key Laboratory of Dark Matter and Space Astronomy, Purple Mountain Observatory, Chinese Academy of Sciences, Nanjing, 210033, People's Republic of China}
\newcommand{\USTC}{School of Astronomy and Space Science, University of Science and Technology of China, Hefei, Anhui 230026, People's Republic of China}

\title{Constraints on the phase transition and nuclear symmetry parameters from PSR $\mathrm{J}0740+6620$ and multimessenger data of other neutron stars}
\author{Shao-Peng Tang}
\author{Jin-Liang Jiang}
\author{Ming-Zhe Han}
\author{Yi-Zhong Fan}
\email[Corresponding author.~]{yzfan@pmo.ac.cn}
\author{Da-Ming Wei}
\affiliation{\PMO}
\affiliation{\USTC}
\date{\today}

\begin{abstract}
Recently, the radius of neutron star (NS) {PSR J0740+6620} was measured by Neutron Star Interior Composition Explorer (NICER) and an updated measurement of neutron skin thickness of ${}^{208}$Pb ($R_{\rm skin}^{208}$) was reported by the PREX-II experiment. These new measurements can help us better understand the unknown equation of state (EOS) of dense matter. In this work, we adopt a hybrid parameterization method, which incorporates the nuclear empirical parameterization and some widely used phenomenological parameterizations, to analyze the results of nuclear experiments and astrophysical observations. With the joint Bayesian analysis of GW170817, PSR J0030+0451, and PSR J0740+6620, the parameters that characterize the ultradense matter EOS are constrained. We find that the slope parameter $L$ is approximately constrained to $70_{-18}^{+21}$ MeV, which predicts $R_{\rm skin}^{208}=0.204^{+0.030}_{-0.026}\,{\rm fm}$ by using the universal relation between $R_{\rm skin}^{208}$ and $L$. The bulk properties of canonical $1.4\,M_\odot$ NS (e.g., $R_{1.4}$ and $\Lambda_{1.4}$) as well as the pressure ($P_{2\rho_{\rm sat}}$) at two times the nuclear saturation density are well constrained by the data; {\it i.e.}, $R_{1.4}$, $\Lambda_{1.4}$, and $P_{2\rho_{\rm sat}}$ are approximately constrained to $12.3\pm0.7$ km, $330_{-100}^{+140}$, and $4.1_{-1.2}^{+1.5}\times10^{34}\,{\rm dyn\,cm^{-2}}$, respectively. Besides, we find that the Bayes evidences of the hybrid star and normal NS assumptions are comparable, which indicates that current observation data are compatible with quarkyonic matter existing in the core of massive star. Finally, in the case of normal NS assumption, we obtain a constraint for the maximum mass of nonrotating NS $M_{\rm TOV}=2.30^{+0.30}_{-0.18}$ $M_\odot$. Based on this result and the current observational and theoretical knowledge about the NS population and its EOS, we find that a binary black hole merger scenario for GW190814 is more plausible. All of the uncertainties reported above are for 68.3\% credible levels.
\end{abstract}
\maketitle

\section{Introduction}
The unknown equation of state (EOS) of dense matter can be constrained from the observations of neutron stars (NSs) that serve as unique astrophysical laboratory for learning the behavior of matter under extreme physical conditions. Steady progress on constraining EOS with the observed maximum mass and traditional mass-radius ($M$-$vs$-$R$) measurements of NSs has been made, and breakthrough was achieved especially for recent years, owing to not only the tidal deformability measurements from the remarkable observations of the binary NS (BNS) merger event GW170817 by advanced LIGO and Virgo detectors \citep{2017PhRvL.119p1101A}, but also the first simultaneous precise mass-radius measurement of the isolated NS {PSR J0030+0451} by Neutron Star Interior Composition Explorer (NICER) \citep{ 2019ApJ...887L..24M, 2019ApJ...887L..21R}. The gravitational wave (GW) imprints from tidal effect provide us a totally new avenue to probe the internal structure of NS, while the mass and radius measured with the novel pulse profile modeling method are reasonably more reliable than traditional spectroscopic measurements. Benefiting from these two observations, the joint analyses of them have set stringent constraints on the EOS \citep{2019ApJ...885...39J, 2020ApJ...892...55J, 2020PhRvD.101l3007L, 2020ApJ...888...12M, 2020ApJ...893L..21R, 2021arXiv210305408H}. 

On the other hand, nuclear experiments and theories also place tight constraints on the relatively low density part of EOS (especially for the symmetry energy and its density dependence), which consistently give an intersection in the space of symmetry energy parameters (see, e.g., Fig.~2 in Ref.~\citep{2020PhRvL.125t2702D}). Recently, the PREX-II experiment updated its result for the neutron-skin thickness of ${}^{208}$Pb, $R_{\rm skin}^{208}=0.283\!\pm\!0.071\,{\rm fm}$ \citep{2021PhRvL.126q2502A}. Using the well-established relation between the thickness $R_{\rm skin}^{208}$ and slope parameter $L$ (see, e.g., Ref.~\citep{2001PhRvC..64b7302T,2014EPJA...50...27V,2016PhRvC..93e1303R}), this measurement, however, indicates a rather high $L$ value \citep{2021PhRvL.126q2503R}, incompatible with other determinations and thus, challenges our understanding of nuclear matter.

For the first time, a radius measurement was announced by the NICER team for the millisecond pulsar {PSR J0740+6620} that with the highest mass known, which allows us to probe the EOS at densities much higher than those based on previous NS observations. Informed by the radio timing \citep{2020NatAs...4...72C, 2021ApJ...915L..12F} and {\it XMM-Newton} spectroscopy, the inferred radius of this massive NS is constrained to $12.39_{-0.98}^{+1.30}\,{\rm km}$ by \citet{2021arXiv210506980R}, and $13.7_{-1.5}^{+2.6}\,{\rm km}$ by \citet{2021arXiv210506979M}, at 68\% credible level. Though {PSR J0740+6620} is much heavier than {PSR J0030+0451} (they differ in mass by $>50\%$), they almost share similar radius. This result rules out many theoretical models that predict very ``squishy" stars and instead favors a much stiffer EOS. However, previous results based on sole GW data suggest that ``soft'' EOSs, which predict small tidal deformability, are favored over ``stiff'' EOSs \citep{2018PhRvL.121p1101A, 2018PhRvL.121i1102D}. 

Interestingly, the posterior of the combined tidal parameter ($\tilde{\Lambda}$) of GW170817 presents a bimodal distribution, in which the second peak is favored if we further include other measurements to perform joint analysis \citep{2021PhRvD.103f3026T}. Meanwhile, the NS nature of the secondary object in GW190814 is inconsistent with either the $M_{\rm TOV}$ (maximum mass of nonrotating NS) determinations \citep{2021ApJ...908L..28N} by the multimessenger analyses of GW170817/GRB 170817A/ AT2017gfo \citep{2018ApJ...852L..25R, 2018PhRvD..97b1501R, 2019PhRvD.100b3015S, 2020PhRvD.101f3029S, 2020ApJ...904..119F} or the constraints obtained from energetic heavy-ion collisions \citep{2020PhRvC.102f5805F}, while the black hole (BH) nature also challenges our knowledge about the formation of compact-object binaries \citep{2020ApJ...896L..44A}. These phenomena indicate that the revisit of constraining EOS with the inclusion of {PSR J0740+6620} is necessary for better understanding all of these measurements. Based on the new measurement of $R_{\rm skin}^{208}$ and the new observation of {PSR J0740+6620}, various works have been done \citep{2021arXiv210505132A, 2021arXiv210502886B, 2021arXiv210210074E, 2021Univ....7..182L, 2021arXiv210508688P, 2021arXiv210506981R, 2021arXiv210205267Y, 2021arXiv210511031Z}.

In our previous work \citep{2021PhRvD.103f3026T}, we have proposed a hybrid parameterization method to construct a generic phenomenological EOS model that is flexible to resemble various theoretical EOSs. In this work, we improve this method by incorporating the parabolic expansion-based nuclear empirical parameterization around the nuclear saturation density, which is similar to the type of models used in Refs.~\citep{2010ApJ...722...33S, 2021PhRvD.103j3015B}. The coefficients of the expansion, known as the nuclear empirical parameters, can be conveniently related to nuclear experiments, theories, and astrophysical observations. Then, we apply our model to the joint Bayesian analysis of GW data and NICER's measurements. As a result, the parameters describing the EOS are constrained. With the reconstruction of posterior samples, we update the credible region of EOS, the $M$-$vs$-$R$ relation, the bulk properties of the canonical $1.4\,M_\odot$ NS, and the pressure at around $2\,\rho_{\rm sat}$ ($\rho_{\rm sat}$ means the nuclear saturation density). We also reevaluate the Bayes factor between the hybrid star and normal NS assumptions. A comparison is made between the result of PREX-II experiment and the prediction of $R_{\rm skin}^{208}$ translated from the inferred slope parameter $L$ using the universal relation from \citet{2014EPJA...50...27V}. Finally, the nature of the secondary object of GW190814 is discussed based on current observational and theoretical knowledge of the NS population and its EOS.

This work is organized as follows: The parametrized EOS models, priors, observation data, and Bayesian inference method are described in Sec.~\ref{sec:methods}. Our main results are presented in Sec.~\ref{sec:results}, and the conclusion is summarized in Sec.~\ref{sec:summary}. Throughout this work, the uncertainties are for a 68.3\% confidence level unless specifically noticed.

\section{Methods}\label{sec:methods}
In our previous work \citep{2021PhRvD.103f3026T}, we have divided the EOS into five segments, in which a combination of three widely used phenomenological parametrization models are implemented, {\it i.e.}, piecewise polytrope \citep{2009PhRvD..79l4032R, 2009PhRvD..80j3003O}, causal spectral representation \citep{2018PhRvD..97l3019L}, and constant-speed-of-sound (CSS) parametrization \citep{2013PhRvD..88h3013A}. In this work, we replace SLy \citep{2001A&A...380..151D} EOS in the first segment with BPS \citep{1971NuPhA.175..225B} and NV \citep{1973NuPhA.207..298N} EOSs, and a polytrope in the second segment with a schematic expression representing charge-neutral uniform baryonic matter in $\beta$ equilibrium.

We know, for degenerate relativistic electrons, the chemical potential and energy density of noninteracting Fermi gas model are given by
\begin{equation}
\begin{aligned}
\mu_{\rm e}(\rho, x) &= \hbar c\left(\frac{3\pi^2\rho x}{m_{\rm N}}\right)^{1/3}, \\
\varepsilon_{\rm e}(\rho, x) &= \frac{3}{4}\mu_{\rm e}(\rho, x)\frac{\rho x }{m_{\rm N}},
\end{aligned}
\end{equation}
where $\rho$ is the nucleon mass density, $m_{\rm N}$ is the average rest mass of nucleons, and $x$ is the proton fraction.
Meanwhile, in most theoretical models of cold uniform nuclear matter, the energy at a given density can be well approximated by the standard quadratic expansion,
\begin{equation}
\frac{E_{\rm nuc}}{A}(\rho, x)\simeq\frac{E_{\rm SNM}}{A}(\rho)+S_2(\rho)(1-2x)^2.
\end{equation}
The energy per nucleon in symmetric nuclear matter is described as
\begin{equation}
\frac{E_{\rm SNM}}{A}(\rho)\simeq E_0(\rho_{\rm sat})+\frac{K_0}{18}\left(\frac{\rho-\rho_{\rm sat}}{\rho_{\rm sat}}\right)^2,
\end{equation}
where $E_0(\rho_{\rm sat})=-15.9{\rm MeV}$ and $K_0$ denote the incompressibility.
The symmetry energy is
\begin{equation}
S_2(\rho)\simeq S_v+\frac{L}{3}\left(\frac{\rho-\rho_{\rm sat}}{\rho_{\rm sat}}\right)+\frac{K_{\rm sym}}{18}\left(\frac{\rho-\rho_{\rm sat}}{\rho_{\rm sat}}\right)^2,
\end{equation}
where $S_v$, $L$, and $K_{\rm sym}$, respectively, denote the lowest order symmetry energy at $\rho_{\rm sat}$, the slope, and curvature of the symmetry energy. Therefore, within the $npe(\mu)$ model for the outer core of NS, the energy density and total pressure can be approximated by
\begin{equation}
\begin{aligned}
\varepsilon(\rho, x) &\approx \varepsilon_{\rm e}(\rho, x)+\frac{\rho}{m_{\rm N}}[\frac{E_{\rm nuc}}{A}(\rho, x)\\
&\quad+xm_{\rm p}c^2+(1-x)m_{\rm n}c^2]\\
&\approx \varepsilon_{\rm e}(\rho, x)+\frac{\rho}{m_{\rm N}}\left[\frac{E_{\rm nuc}}{A}(\rho, x)+m_{\rm N}c^2\right],\\
p(\rho, x) &= \rho\frac{d\varepsilon(\rho, x)}{d\rho}-\varepsilon(\rho, x)\\
&=\frac{\rho^2}{m_{\rm N}}\frac{d\frac{E_{\rm nuc}}{A}(\rho, x)}{d\rho}+\frac{1}{4}\frac{\rho x}{m_{\rm N}}\mu_{\rm e}.
\end{aligned}
\end{equation}
Then the adiabatic index is derived to
\begin{equation}
\begin{aligned}
\Gamma_{\rm nuc}(\rho, x) &= \frac{\varepsilon+p}{p}c_s^2(\rho, x)\\
&=2+\frac{1}{p}\left[\frac{\rho^3}{m_{\rm N}}\frac{\partial^2 \frac{E_{\rm nuc}}{A}(\rho, x)}{\partial \rho^2}-\frac{1}{6}\frac{\rho x}{m_{\rm N}}\mu_{\rm e}\right],
\end{aligned}
\end{equation}
where
\begin{equation}
\frac{\partial^2 \frac{E_{\rm nuc}}{A}(\rho, x)}{\partial \rho^2}=\frac{K_0+K_{\rm sym}(1-2x)^2}{9\rho_{\rm sat}^2}.
\end{equation}
Neutron star matter in $\beta$ equilibrium satisfies the condition that the total energy is at a minimum with respect to its composition; then, we have
\begin{equation}
\begin{aligned}
0&=\frac{\partial \varepsilon(\rho, x)/n}{\partial x}\\
&=-4S_2(\rho)(1-2x)+(m_{\rm p}-m_{\rm n})c^2+\mu_{\rm e}.
\end{aligned}
\end{equation}
Thus, the proton fraction can be determined at a given density, with which both the energy density, the pressure, and adiabatic index are determined as well. Besides, the above parabolic expansion is jointed to the inner crust at the position of crust-core transition (see also, e.g., Ref.~\citep{2018ApJ...859...90Z}), which is approximately determined by the vanishing effective incompressibility of $npe(\mu)$ matter at $\beta$ equilibrium under the charge neutrality condition \citep{2007PhRvC..76b5801K,2007PhR...442..109L}; {\it i.e.},
\begin{equation}
\begin{aligned}
K_\mu &= \rho^2\frac{\partial^2 \frac{E_{\rm nuc}}{A}(\rho, x)}{\partial \rho^2}+2\rho\frac{\partial \frac{E_{\rm nuc}}{A}(\rho, x)}{\partial \rho}\\
&\quad-\frac{2(1-2x)^2}{S_2(\rho)} \left(\rho\frac{\partial S_2(\rho)}{\partial \rho}\right)^2=0.
\end{aligned}
\end{equation}
We also use
\begin{equation}\label{eq:rhoguess}
\rho_{\rm t}^{\rm init}=\left(\frac{866}{1621+K_{\rm sym}/{\rm MeV}-5L/{\rm MeV}}\right)^{2}\rho_{\rm sat}
\end{equation}
as an initial guess value\footnote{In practice, we solve $K_\mu=0$ with two root-finding algorithms. The first is the Steffenson method, which combines the basic Newton algorithm with an Aitken``delta-squared” acceleration. When this method is failed to converge (a bad initial guess value), we then use a more robust but slower Brent-Dekker method to find the root. Equation~(\ref{eq:rhoguess}) is a fitting result of crust-core transition density as a function of $K_{\rm sym}$ and $L$, which largely increases the efficiency for using Steffenson method and hence, accelerates our codes.} to accelerate the solution of $K_\mu=0$. 

The adiabatic indices used in the other three segments are identical to those in Ref.~\citep{2021PhRvD.103f3026T}; {\it i.e.},
\begin{equation}\label{eq:Gamma}
    \Gamma(\varepsilon,p,h) = \begin{cases}
                        \Gamma_{\rm crust} & \quad \rho<\rho_0, \\
                        \Gamma_{\rm nuc}(\rho, x) & \quad \rho_0<\rho \leq \rho_1, \\
                        \frac{1}{1+\Upsilon(h,v_k)}\frac{\varepsilon+p}{p} & \quad \rho_1<\rho \leq \rho_2, \\
                        \Gamma_{\rm m} & \quad \rho_2<\rho \leq \rho_2\!+\!\Delta \rho, \\
                        c_{\rm q}^2\frac{\varepsilon+p}{p} & \quad \rho>\rho_2\!+\!\Delta \rho,
                    \end{cases}
\end{equation}
where $\varepsilon$, $p$, $h$, and $\rho$ denote, respectively, the internal energy density (including the rest mass contribution), the total pressure, the pseudo enthalpy defined by $h(p)=\int_0^p\mathop{}\!\mathrm{d}p^\prime/[\varepsilon(p^\prime)+p^\prime]$, and the rest-mass density that can be calculated by $\rho=(\varepsilon+p)/\exp{\!(h)}$. The nuclear empirical expansion is only valid within $\rho_0<\rho \leq \rho_1$; here, $\rho_0$ is the crust-core transition density determined by $K_\mu=0$. Since different choices of $\rho_1$ give almost the same Bayes evidence (see Ref.~\citep{2021PhRvD.103j3015B}) and the results of previous study \citep{2010ApJ...722...33S} indicate that $\rho_1 \simeq1.8\,\rho_{\rm sat}$ is plausible, therefore, we fix $\rho_1$ to $1.85\,\rho_{\rm sat}$ without loss of generality. Meanwhile, the parabolic approximation should be appropriate in this work because we only use it up to about $1.85\,\rho_{\rm sat}$ \citep{2009PhRvC..80a4322C}. The expression of $\Upsilon(h,v_k)$ is
\begin{equation}\label{eq:Upsilon}
    \Upsilon(h,v_k)=\exp \left\{ \sum_{k=0}^3 v_k \left[ \log{\left( \frac{h}{h_{\rm ref}}\right)} \right]^k \right\},
\end{equation}
where $v_k$ are the expansion coefficients, and $h_{\rm ref}$ is the pseudoenthalpy at the density of $\rho_1$. Therefore, the EOS can be described by 12 free parameters, {\it i.e.}, $\vec{\theta}_{\rm EOS}=\{K_0, S_v, L, K_{\rm sym}, v_0, v_1, v_2, v_3, \rho_2, \Delta \rho, \Gamma_{\rm m}, c_{\rm q}^2\}$, where $\rho_2$ means the dividing density, $\Gamma_{\rm m}$ is the adiabatic index within the density jump $\Delta \rho$, and $c_{\rm q}^2$ is the sound velocity parameter that describes the EOS with CSS parametrization above $\rho_2\!+\!\Delta \rho$. Each set of parameters can be translated to a possible EOS and uniquely determine the macroscopic relations like $M$-$vs$-$R$ curve. Thus, we can inversely constrain these parameters with a series of accurate observations within the Bayesian framework.

We empirically construct the model of phase transition (PT) taking place in NS with $\Gamma_{\rm m} \in[0.01, 1.4]$ and the range of $[1.4, 10]$ otherwise [{\it i.e.}, there is no phase transition (NPT)]. We choose the ranges for other parameters of $\vec{\theta}_{\rm EOS}$ with $K_0\in[210, 270]\,{\rm MeV}$, $S_v\in[28, 44]\,{\rm MeV}$, $L\in[40, 120]\,{\rm MeV}$, $K_{\rm sym}\in[-400, 100]\,{\rm MeV}$, $v_0 \in[-1.2, 3.6]$, $v_1 \in[-5.3, 2.8]$, $v_2 \in[-5.2, 7.3]$, $v_2 \in[-4.1, 1.8]$, $\rho_2 \in[1.85, 5]\,\rho_{\rm sat}$, $\Delta \rho \in[0.01, 3.0]\,\rho_{\rm sat}$, and $c_{\rm q}^2 \in[1/3, 1]$\footnote{This range is empirically designed to ensure that our PT and NPT models are not mixed or contaminated with each other. For example, in the NPT model, if we allow $c_{\rm q}^2$ to vary between 0 and 1, we may have a small $c_{\rm q}^2$, which will masquerade the PT model. Therefore, it is difficult for our phenomenological model to simultaneously consider a wider prior of $c_{\rm q}^2$ and perform Bayesian model selection of PT/NPT models. There may be only one situation (see middle panel of Fig.~2 of \citet{2020arXiv201110940K}) that our models do not include; {\it i.e.}, after phase transition, there is a very low value of $c_{\rm q}^2$. However, this scenario is not favored as suggested by various works (see, e.g., \citep{2015PhRvL.114c1103B, 2021PhRvD.104f3003L})}, where the ranges of parameters $v_k$ are determined by fitting the theoretical EOSs with the causal spectral representation method. Additionally, all of the parametrized EOSs satisfy the following conditions:
\begin{enumerate}
\item Causality constraint and thermal stability,
\item $\Gamma \in [1.4, 10]$ when extending the causal spectral representation to $5\rho_{\rm sat}$,
\item Maximum central density of nonrotating NS should exceed $\rho_2$ for PT model,
\item The initial guess value of crust-core transition density $\rho_{\rm t}^{\rm init}<\rho_{\rm sat}$,
\item Maximum mass limits $M_{\rm TOV}\in[2.0, 2.9]\,M_\odot$\footnote{The left boundary is chosen based on the observed massive NSs.}.
\end{enumerate}

In this work, our data set $\mathcal{D}$ includes observation data of GW170817 \citep{2017PhRvL.119p1101A}, mass-radius measurements of {PSR J0030+0451} \citep{2019ApJ...887L..21R, 2019ApJ...887L..24M}, and {PSR J0740+6620}\footnote{We take the data file ``STU/NICERxXMM/FI\_H/run10" that includes the information of {\it XMM-Newton} from \url{https://zenodo.org/record/4697625\#.YKMcuy0tZQJ}.} \citep{2021arXiv210506980R, 2021arXiv210506979M}. Assuming that compact stars share the same EOS, we take the likelihood,
\begin{equation}\label{eq:likelihood}
    \mathcal{L}=\mathcal{L}_{\rm GW}(d\mid\vec{\theta}_{\rm GW})\times \prod_{i} \mathcal{P}_i(M(\vec{\theta}_{\rm EOS}, h_i), R(\vec{\theta}_{\rm EOS}, h_i))
\end{equation}
to constrain the parameters $\vec{\theta}_{\rm EOS}$ that characterize the ultradense matter EOS by performing Bayesian inference with {\sc Bilby} \citep{2019ascl.soft01011A} and {\sc dynesty} \citep{2020MNRAS.493.3132S} packages. 
For the mass-radius observations of {PSR J0030+0451} and {PSR J0740+6620}, we use the posterior samples ($\vec{S}$) to construct the kernel density estimate (KDE) as $\mathcal{P}_i(M,R)={\rm KDE}(M,R\mid \vec{S})$. Each pair of $(M,R)$ are calculated by varying the central enthalpy $h_i$ in the range of [0.06, 1.0]. For the GW likelihood $\mathcal{L}_{\rm GW}$, we calculate it using the random forest interpolator \citep{2020MNRAS.499.5972H} with four parameters $\vec{\theta}_{\rm GW} = \{\Lambda_1(m_1^{\rm src},\vec{\theta}_{\rm EOS}),\Lambda_2(m_2^{\rm src},\vec{\theta}_{\rm EOS}), \mathcal{M}_{\rm c}, q\}$, where $\Lambda_{1,2}$ are dimensionless tidal deformabilities that are mapped from source frame masses $m_{1,2}^{\rm src}$ using EOS parameters. All of the parameters ($\vec{\theta}_{\rm EOS}$, $h_i$, $\mathcal{M}_{\rm c}$, and $q$) are uniformly distributed in their domains.

\section{Results}\label{sec:results}
\begin{figure*}
    \centering\vspace{-6mm}
    \includegraphics[width=0.75\textwidth]{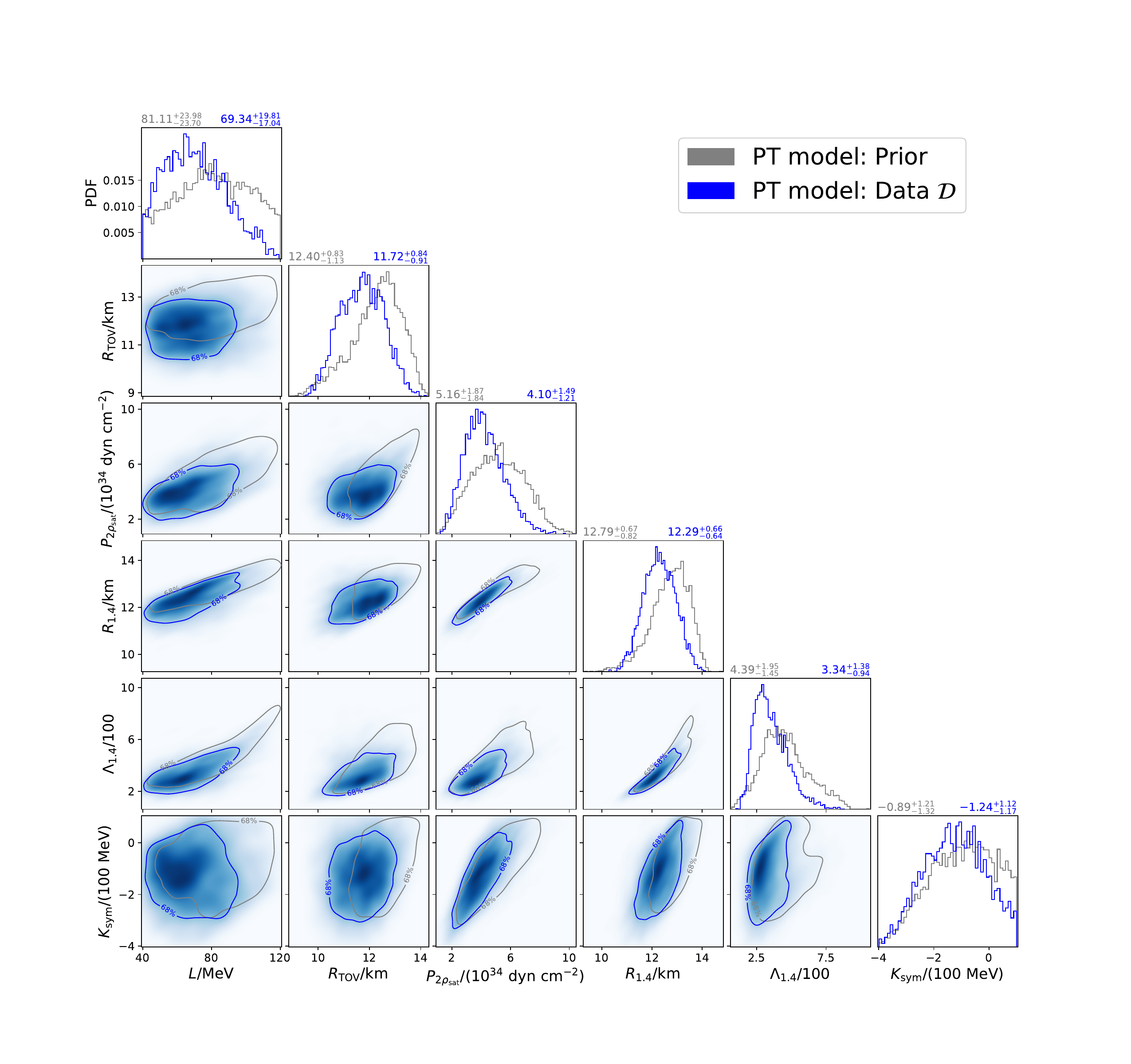}\vspace{-8mm}
    \includegraphics[width=0.75\textwidth]{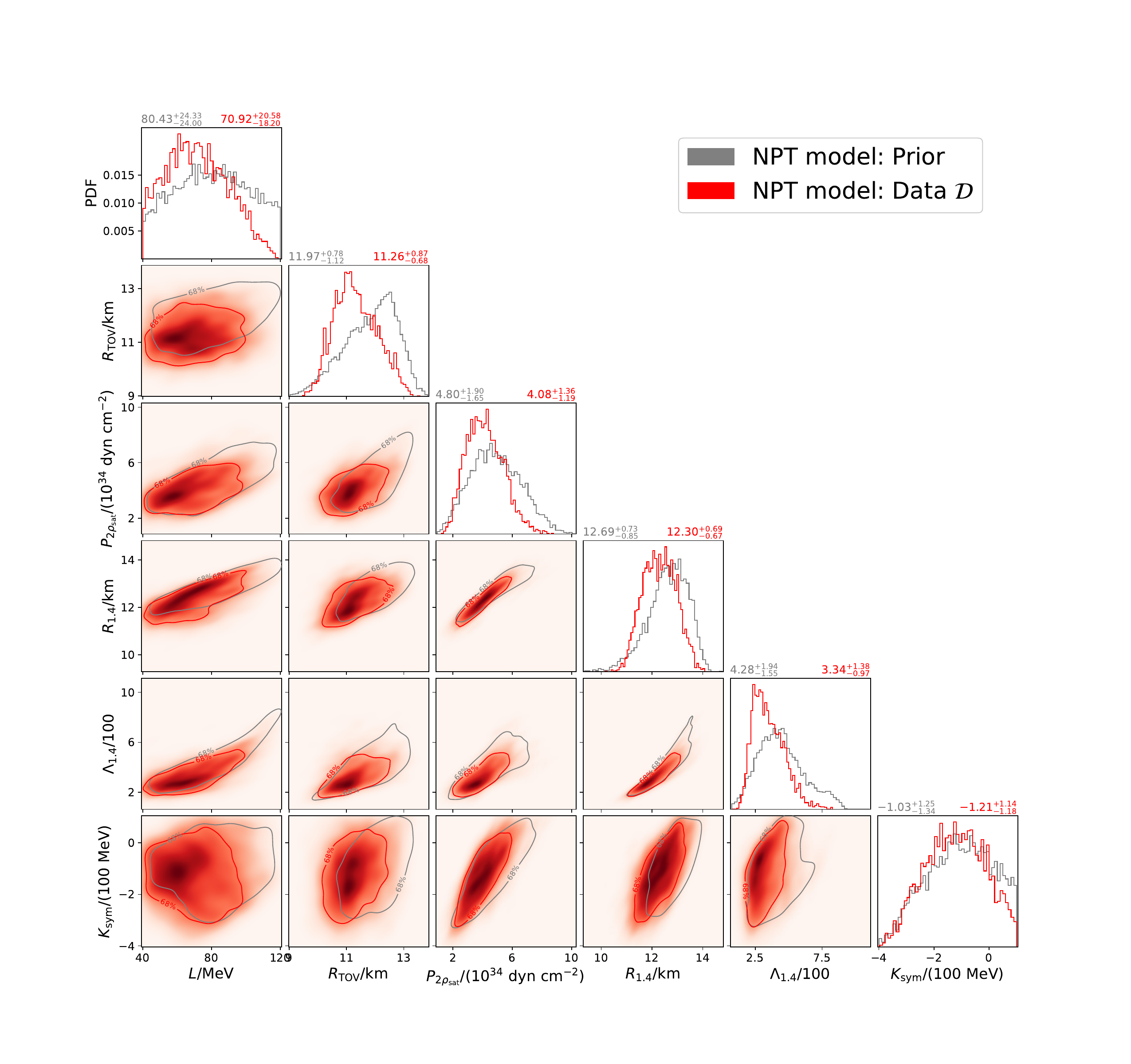}\vspace{-8mm}
    \caption{Corner plots of symmetry energy parameters ($L$, $K_{\rm sym}$), bulk properties of canonical $1.4\,M_\odot$ NS ($R_{1.4}$, $\Lambda_{1.4}$), pressure at $2\,\rho_{\rm sat}$, and radius corresponding to $M_{\rm TOV}$. Priors (considering the enforced conditions) are represented by gray color, and the top and bottom panels respectively show the results of PT and NPT models. Meanwhile, values in the diagonal plots are 68.3\% credible intervals (Fig.~\ref{fig:rho2} and Fig.~\ref{fig:vk} are the same).}
    \label{fig:LKsym}
    \hfill
\end{figure*}
Here, we only report the results based on the measurements from \citet{2019ApJ...887L..21R, 2021arXiv210506980R}, since the mass-radius measurements of both {PSR J0030+0451} and {PSR J0740+6620} given by the two NICER groups are consistent considering the current measurement errors. We expect that different combinations of these measurements will not significantly change our results.

The posterior distributions of the empirical parameters $L$ and $K_{\rm sym}$, the pressure at $2\,\rho_{\rm sat}$, the bulk properties of $1.4\,M_\odot$ NS, {\it i.e.}, $R_{1.4}$ and $\Lambda_{1.4}$, and the radius of NS at maximum mass configuration, are presented in Fig.~\ref{fig:LKsym}. We can see that for both models, the radius and tidal deformability of a canonical $1.4\,M_\odot$ NS are very similar, {\it i.e.}, $12.29^{+0.66}_{-0.64}$ ($12.30^{+0.69}_{-0.67}$) km and $334_{-94}^{+138}$ ($334_{-97}^{+138}$) for PT (NPT) model. The radius corresponding to the $M_{\rm TOV}$ shows a little bit difference, {\it i.e.}, $11.72^{+0.84}_{-0.91}$ ($11.26^{+0.87}_{-0.68}$) km for PT (NPT) model. The slope parameter $L$ is well constrained to $69_{-17}^{+20}$ ($71_{-18}^{+21}$) MeV for PT (NPT) model. We find that strong correlations present between the pressure at $2\,\rho_{\rm sat}$ and bulk property like $R_{1.4}$ (or $\Lambda_{1.4}$) \citep{2016PhR...621..127L,2018PhRvC..98c5804M}. Meanwhile, the radius of $1.4\,M_\odot$ NS is also correlated with the slope parameter $L$, though it is not as strong as analyses based on a specific class of EOS model (e.g., Ref.~\citep{2020ARNPS..70...21Y}). Interestingly, the NS radii for two distinct masses, {\it i.e.}, $R_{\rm TOV}$ and $R_{1.4}$, exhibit a slightly positive correlation. The symmetry incompressibility $K_{\rm sym}$, though is not constrained as well as other parameters by the astrophysical data, favors negative values. As for other nuclear empirical parameters, $K_0$ and $S_v$, current observation data are not informative enough to place any constraints on them (as shown in Fig.~\ref{fig:vk}).

\begin{figure*}
    \centering
    \includegraphics[width=0.75\textwidth]{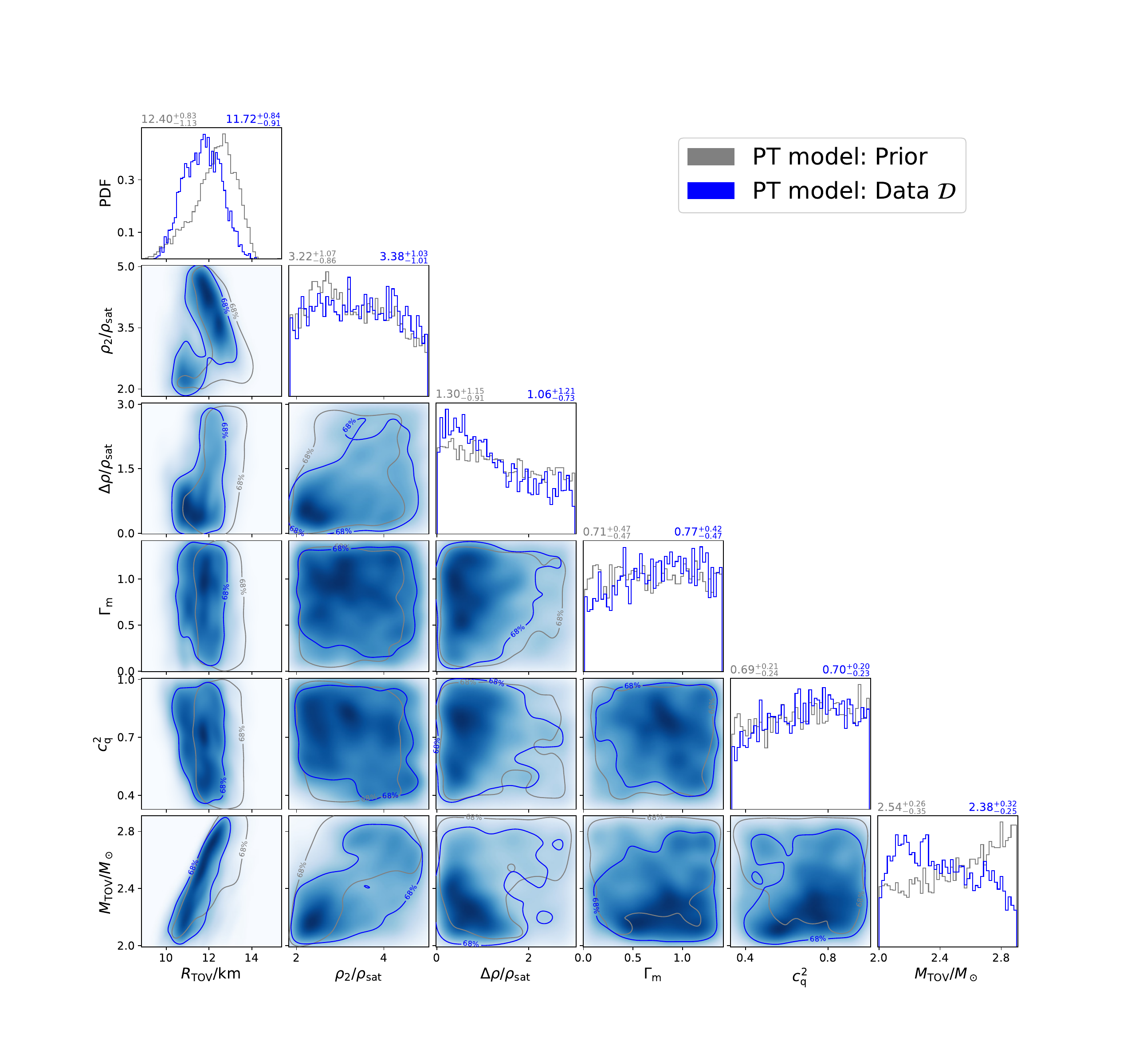}\vspace{-8mm}
    \includegraphics[width=0.75\textwidth]{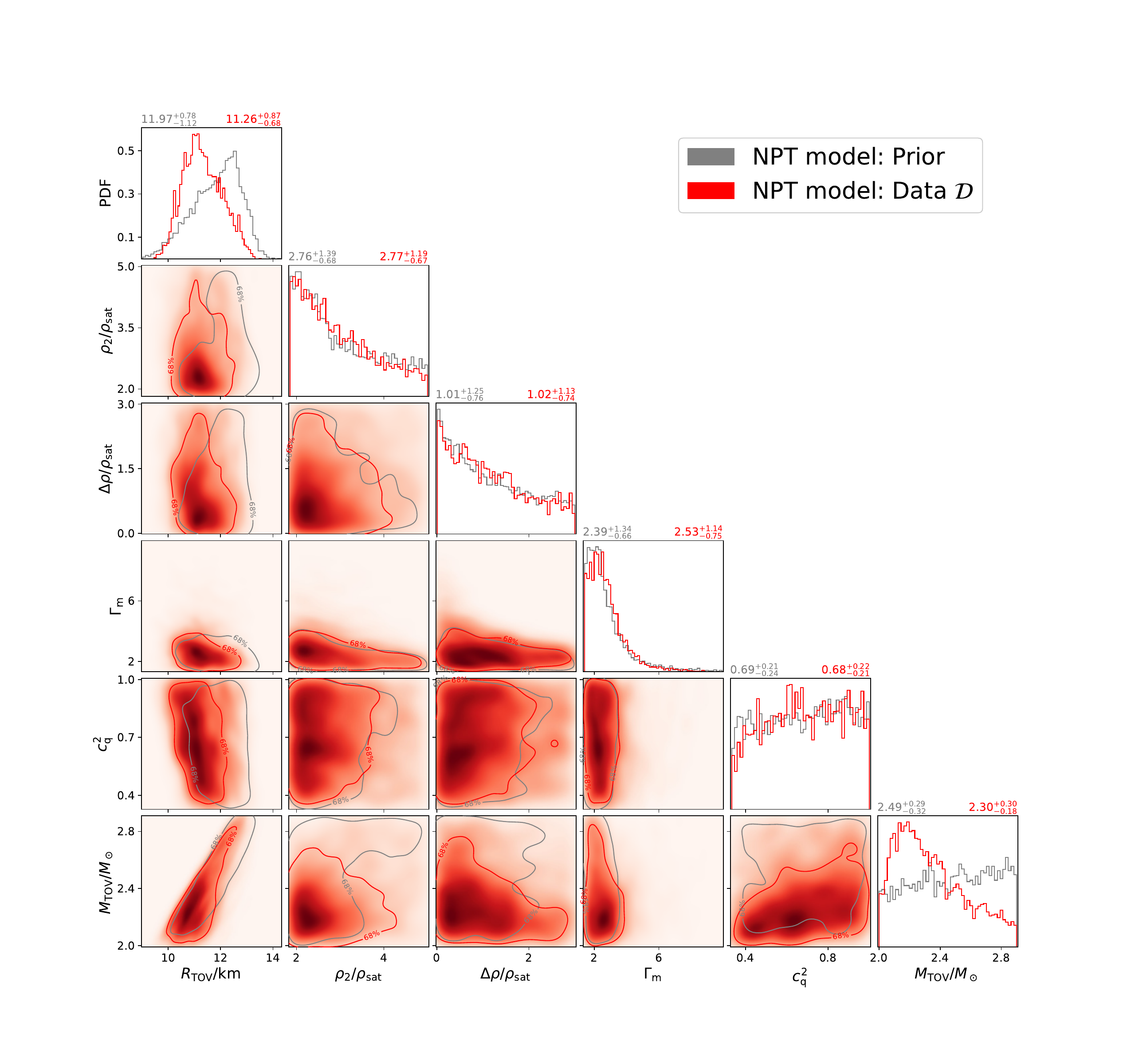}\vspace{-8mm}
    \caption{Corner plots of mass-radius properties at maximum mass configuration and EOS parameters $\{\rho_2, \Delta \rho, \Gamma_{\rm m}, c_{\rm q}^2\}$.}
    \label{fig:rho2}
    \hfill
\end{figure*}
Most of the parameters that determine the EOS at densities exceeding $\rho_{1}$ are less constrained by the observation data $\mathcal{D}$, and the resulting distributions mainly come from the enforced conditions. The speed of sound parameter in both cases and the adiabatic index $\Gamma_{\rm m}$ in the PT case are loosely constrained, while the $\Gamma_{\rm m}$ in the NPT case is tightly constrained by the causality condition. The PT case disfavors the low transition density with a strong phase transition, which mostly results from the maximum mass limit and the radius measurements, while in the NPT case, the region with large $\rho_{2}$ as well as large $\Delta\rho$ is excluded by the causality constraint. As shown in Fig.~\ref{fig:vk}, the spectral parameters are loosely constrained except for $v_{0}$, which is larger than zero with a high probability. Besides, there exists a correlation in all neighboring coefficients (e.g., the $v_1$-$vs$-$v_2$ correlation), which is an inevitable result of the expansion form of the spectral representation.

\begin{figure}
    \centering
    \includegraphics[width=0.98\columnwidth]{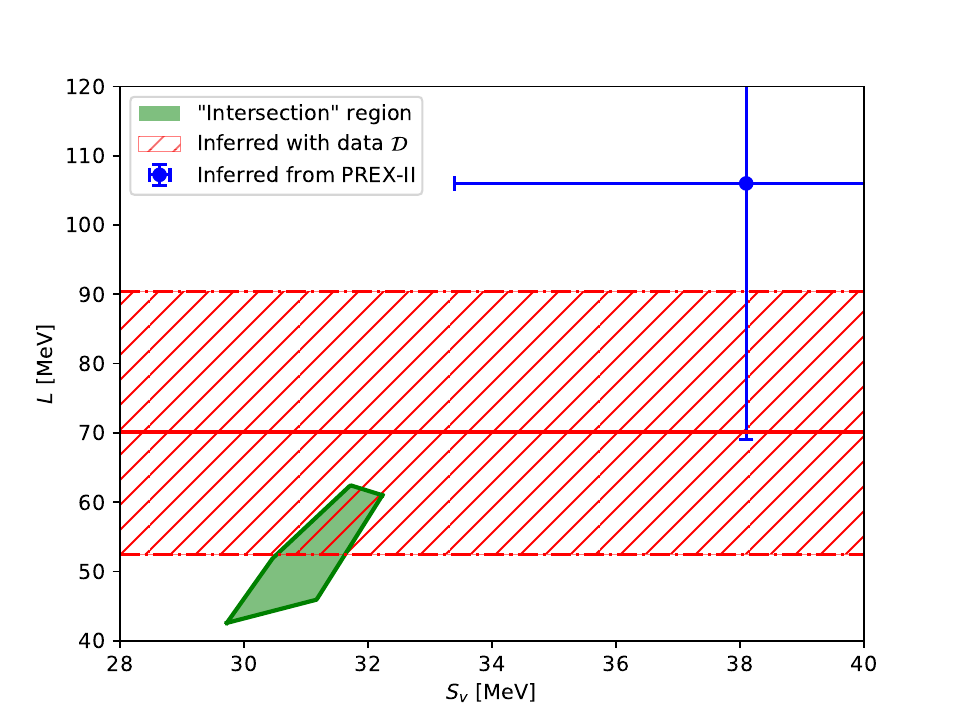}
    \includegraphics[width=0.98\columnwidth]{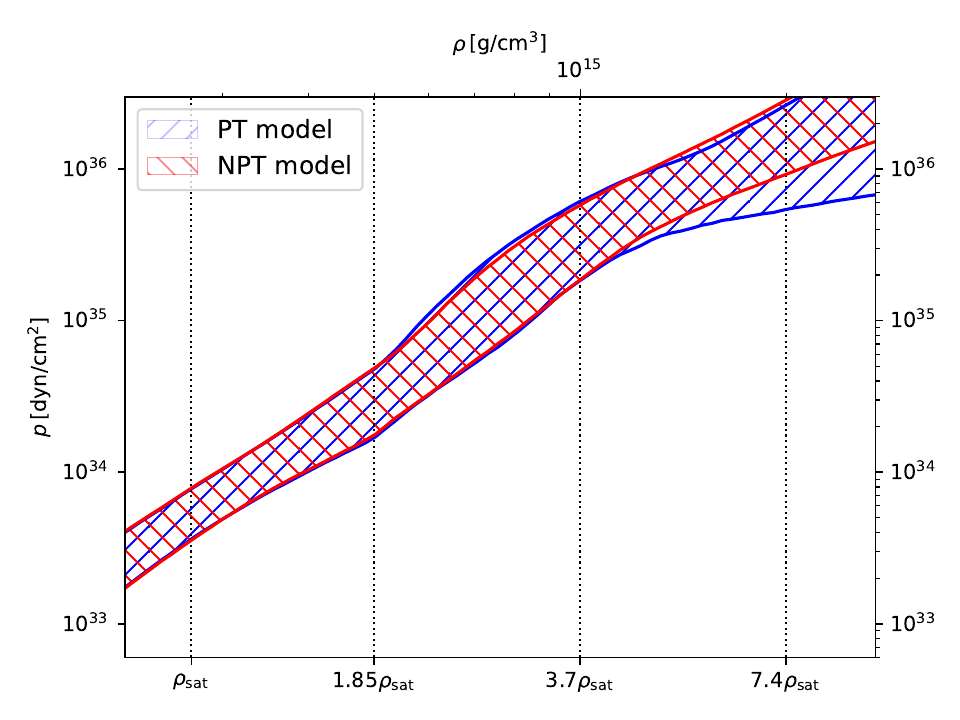}
    \caption{Top panel: a comparison of symmetry energy parameters among the constraints from astrophysical observations (red hatched, $1\sigma$ confidence level), ``intersection" parameter space (green area, $2\sigma$ confidence level) from nuclear experiments \citep{2020PhRvL.125t2702D}, and that inferred from PREX-II (blue errorbar, $1\sigma$ confidence level) \citep{2021PhRvL.126q2503R}. Bottom panel: the 90\% uncertainty regions of constrained EOS in the form of rest-mass density versus the total pressure for PT (blue hatched) and NPT (red hatched) models.}
    \label{fig:prhoSL}
    \hfill
\end{figure}
\begin{figure}
    \centering\vspace{-6mm}
    \includegraphics[width=0.98\columnwidth]{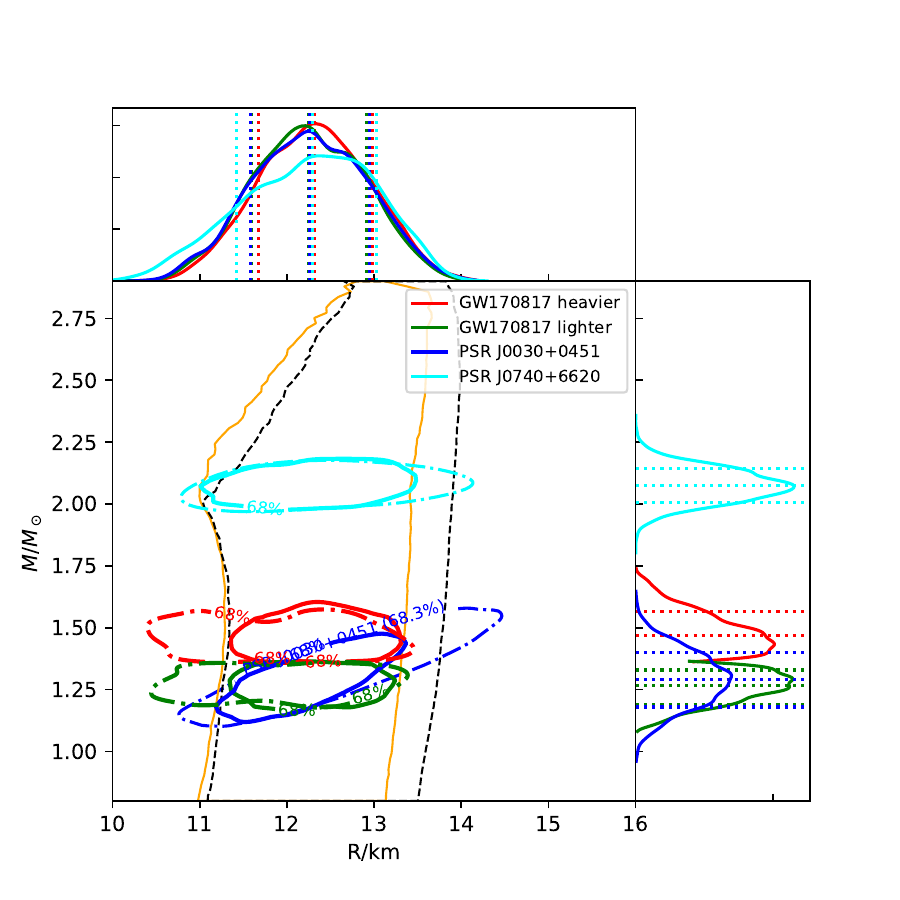}\vspace{-3mm}
    \includegraphics[width=0.98\columnwidth]{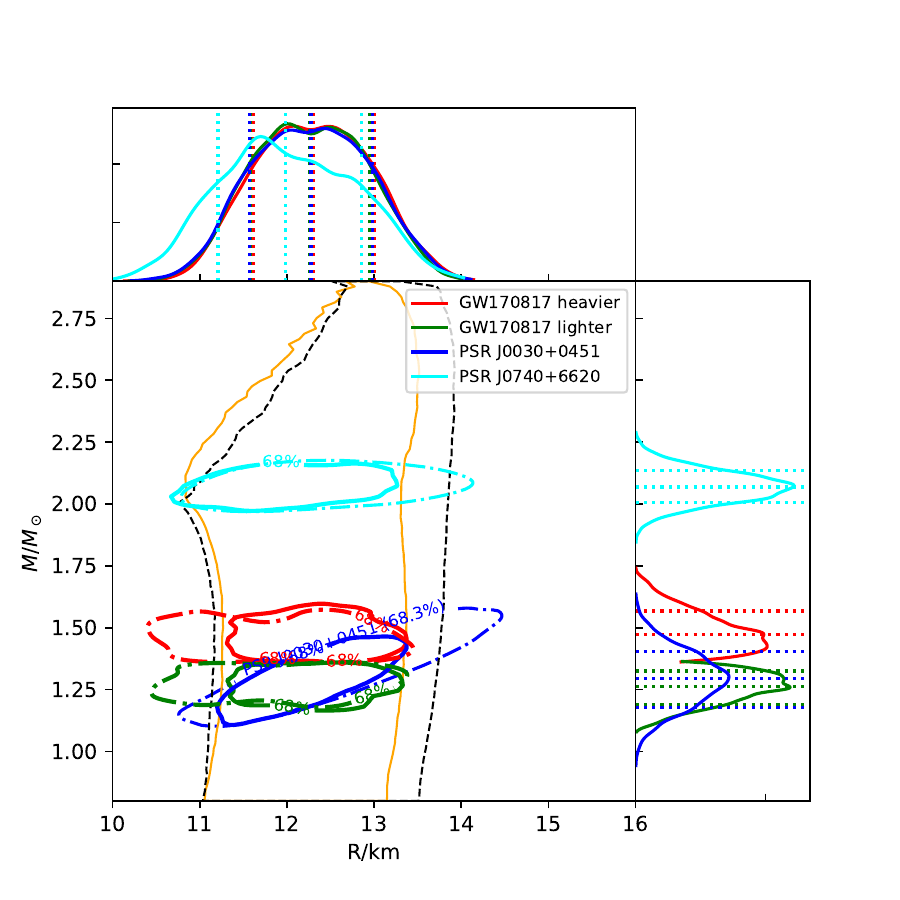}\vspace{-4mm}
    \caption{Posterior $M$-versus-$R$ distributions for PT (top panel) and NPT (bottom panel) models. The black dashed line and orange solid line denote, respectively, the 68.3\% uncertainty region of $M$-versus-$R$ relations of the prior and the posterior obtained with data set $\mathcal{D}$. The $M$-versus-$R$ measurements of {PSR J0030+0451} and {PSR J0740+6620} are represented by the blue and cyan dot-dashed contours, respectively. The red and green dot-dashed contours represent the $M$-versus-$R$ posteriors in the right panel of Fig.~3 of Ref.~\citep{2018PhRvL.121p1101A}. The associated reconstructed $M$-versus-$R$ of these sources are represented by the colored solid contours.}
    \label{fig:mrfit}
    \hfill
\end{figure}
A comparison for constrained symmetry energy parameters is displayed in the top panel of Fig.~\ref{fig:prhoSL}, where the error bar represents the inferred values of $S_v$-vs-$L$ based on PREX-II measurement from Ref.~\citep{2021PhRvL.126q2503R}, the green region represents the consistent determinations by many nuclear experiments and theories \citep{2017ApJ...848..105T,2020PhRvL.125t2702D}, and the red hatched area represents our constraint based on astrophysical data. We report the 90\% regions of constrained rest-mass density and total pressure ($\rho$-$vs$-$p$) relation for PT and NPT models in the bottom panel of Fig.~\ref{fig:prhoSL}, in which both models give rather similar constraints when $\rho\lesssim4\,\rho_{\rm sat}$. Discrepancy in the higher region is the nature result of the models where the NPT model gives stiffer EOSs. As for the sound velocity property of dense matter, current data are still hard to give insight into it and remains less constrained compared with the $\rho$-versus-$p$ relation. The PT and NPT models also give highly consistent mass-radius relation (see Fig.~\ref{fig:mrfit}) and comparable Bayes evidences, with a Bayes factor of $\mathcal{B}^{\rm PT}_{\rm NPT}=0.9$, showing that data nowadays are consistent with having quark matter core in NS, which is similar to the results of Refs.~\citep{2021arXiv210413612S,2021arXiv210505132A}. These phenomena show that the observation properties of Gibbs construction of our PT model can be masqueraded by NPT model. Besides, the joint analysis indicates that NSs in the mass range $[0.8, 2.25]\,M_{\odot}$ have almost the same radius, {\it i.e.}, $[11, 13]\,{\rm km}$, and favor the second peak of the distribution of the radii given by previous results \citep{2018PhRvL.121p1101A}.

\begin{figure}
    \centering
    \includegraphics[width=0.98\columnwidth]{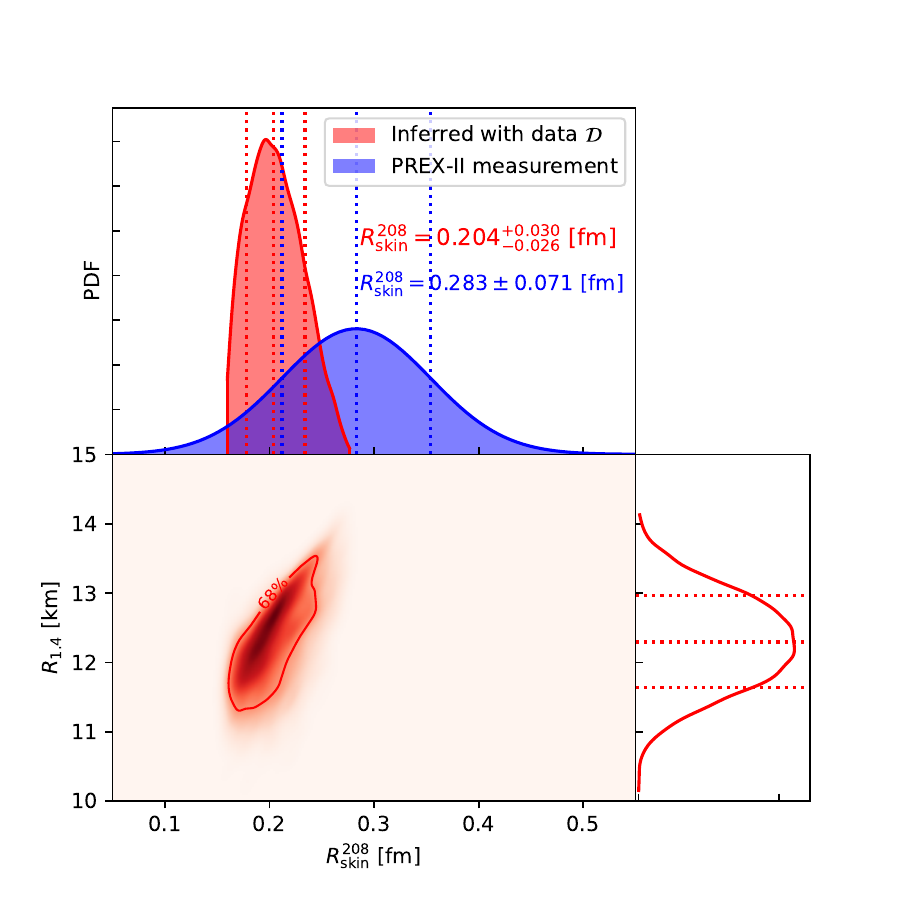}
    \caption{Joint plots of radius of $1.4\,M_\odot$ NS and neutron-skin thickness $R_{\rm skin}^{208}$ transformed from slope parameter $L$ using universal relation in Ref.~\citep{2014EPJA...50...27V}. The blue color represents the measurement from PREX-II.}
    \label{fig:skinr14}
    \hfill
\end{figure}
Shown in the top panel of Fig.~\ref{fig:skinr14} are the joint and marginal plots of neutron skin thickness $R_{\rm skin}^{208}$ and radius of $1.4\,M_\odot$ NS $R_{1.4}$, where $R_{\rm skin}^{208}$ is converted from the slope parameter $L$ using the universal relation $R_{\rm skin}^{208} [{\rm fm}] = 0.101 + 0.00147 \times L [{\rm MeV}]$ \citep{2014EPJA...50...27V}. Note that the distributions of $R_{\rm skin}^{208}$ are consistent between PT and NPT models, so we incorporate an equal number of samples from the posterior of each model to get our overall result (a similar procedure was adopted by \citet{2019PhRvX...9c1040A} to consider the waveform uncertainties). Comparing to the result of PREX-II (with blue color in the panel), astrophysical observations yield a tight constraint on neutron skin thickness, $R_{\rm skin}^{208}=0.204^{+0.030}_{-0.026}\,{\rm fm}$ (at 68.3\% credible level) and tend to favor small value of PREX-II measurement. These results are in good agreement with those of Refs.~\citep{2021arXiv210502886B, 2021arXiv210210074E, 2021Univ....7..182L}. The positive correlation between $R_{\rm skin}^{208}$ and $R_{1.4}$ indicates that the result of PREX-II favors a rather large $R_{1.4}$, which is in tension with the determination of $R_{1.4}$ from the sole GW data of GW170817.

\begin{figure}
    \centering
    \includegraphics[width=0.98\columnwidth]{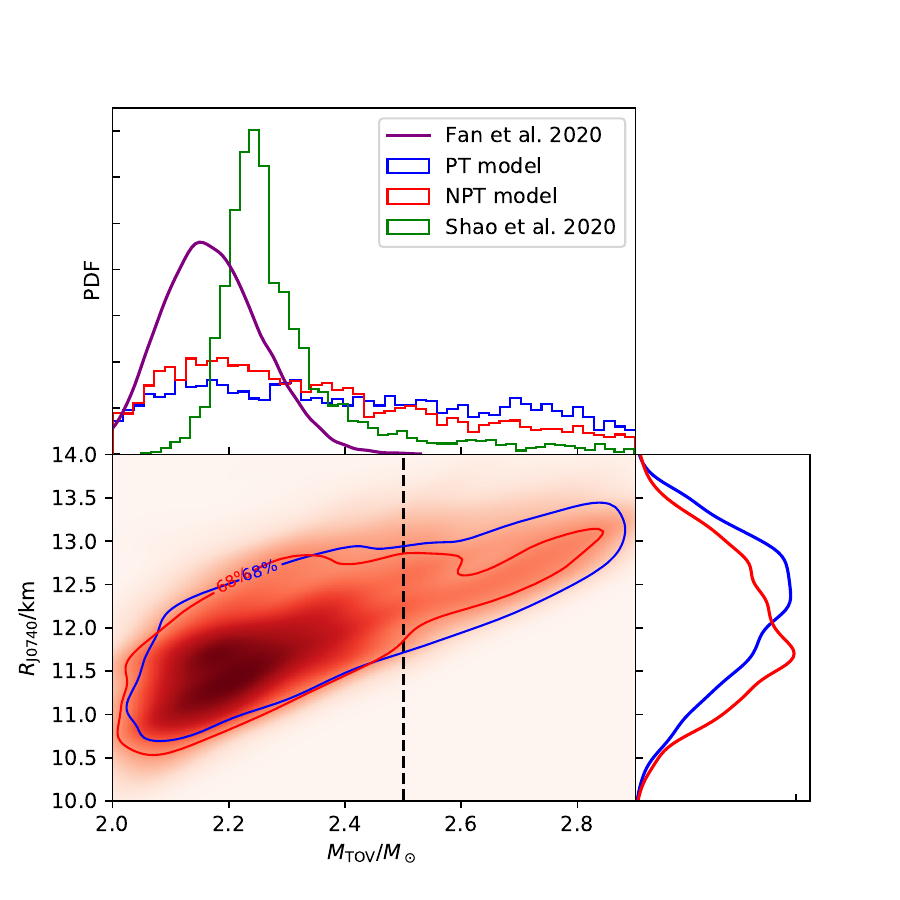}
    \caption{Joint plots of the reconstructed radius of {PSR J0740+6620} from posterior samples and maximum mass of nonrotating NS $M_{\rm TOV}$ for both PT (blue) and NPT (red) models. The purple and green lines represent the probability density distributions of $M_{\rm TOV}$ in Ref.~\citep{2020ApJ...904..119F} and Ref.~\citep{2020PhRvD.102f3006S}, respectively. The dashed vertical line marks the 90\% lower limit of the secondary mass of GW190814. }
    \label{fig:mtov}
    \hfill
\end{figure}
Also, the maximum mass of nonrotating NS is slightly constrained by the data (as shown in Fig.~\ref{fig:rho2}). For the PT (NPT) model, the $M_{\rm TOV}$ is constrained to $2.38^{+0.32}_{-0.25}$ ($2.30^{+0.30}_{-0.18}$) $M_\odot$, at 68.3\% credible level. These results, especially for that of our NPT model, are pretty similar to the results of Ref.~\citep{2021arXiv210508688P}, which gives $M_{\rm TOV}=2.31^{+0.37}_{-0.25}\,M_\odot$ (at 90\% credible interval) when not enforcing the upper bound of $M_{\rm TOV}$ from \citet{2018ApJ...852L..25R}. We notice that the mass and radius at maximum mass configuration reveal a positive correlation; {\it i.e.}, an EOS with larger $M_{\rm TOV}$ may lead to a corresponding larger radius. In Fig.~\ref{fig:mtov}, we show a more interesting joint plot of the reconstructed radius of {PSR J0740+6620} $R_{\rm J0740}$ from the posterior samples and the maximum mass $M_{\rm TOV}$. For both PT and NPT models, $R_{\rm J0740}$ and $M_{\rm TOV}$ also present a positive correlation, which means that large radius of {PSR J0740+6620} will allow NS to support more massive mass. We find that the marginal distributions of $M_{\rm TOV}$ favor, though not strongly, the result of \citet{2020ApJ...904..119F}, which gives $M_{\rm TOV}=2.17_{-0.09}^{+0.09}\,M_\odot$ based on the multimessenger observations of GW170817 and some EOS insensitive relations, and the result of \citet{2020PhRvD.102f3006S}, which gives $M_{\rm TOV}=2.26_{-0.05}^{+0.12}\,M_\odot$ inferred from the population study of Galactic NSs.

These results are helpful in understanding the GW190814 event and revealing the related natures of its secondary component. Due to the large mass ratio of GW190814, the potential tidal effect can not be extracted from its strain data, and the spin effect is also dominated by the massive BH, remaining the dimensionless spin ($\chi$) of the secondary object unconstrained. We know that the effects of fast rotations can increase the maximum mass of NS; for example, the universal relation $M_{\rm crit}=(1+0.0902\mathscr{C}_{\rm TOV}^{-1}\chi^2+0.0193\mathscr{C}_{\rm TOV}^{-2}\chi^4)M_{\rm TOV}$ \citep{2020PhRvD.101f3029S} indicates that a reasonable 20\% enhancement of $M_{\rm TOV}$ can be achieved, if we choose a typical compactness $\mathscr{C}_{\rm TOV}=0.3$ and a dimensionless spin at Kepler rotation $\chi\simeq0.7$. However, an observed maximum dimensionless spin $\chi=0.29$ ({PSR J1959+2048} with MPA1 \citep{1987PhLB..199..469M} EOS; another NS {PSR J1903+0327} has $\chi=0.25$) gives negligible increase to maximum mass (only 2.7\%), hence making the NS-BH nature of GW190814 less possible.

\section{Summary}\label{sec:summary}
We have refined our hybrid parameterization method proposed in Ref.~\citep{2021PhRvD.103f3026T} with the parabolic expansion-based nuclear empirical parameterization in the relatively low density region (e.g., below $1.85\,\rho_{\rm sat}$, which makes the parabolic approximation valid for modeling the EOSs in this work). Based on this method, we empirically constructed PT and NPT models, with which Bayesian analyses were performed using the new measurements of {PSR J0740+6620}, simultaneous $M$-versus-$R$ determination of {PSR J0030+0451} and remarkable observations of GW170817. We find that both PT and NPT models are compatible for explaining current observation data; in other words, massive star with quarkyonic matter in the core is possible. The bulk properties of canonical $1.4\,M_\odot$ NS are well constrained by the data since the masses of three sources we used are close to $1.4\,M_\odot$ and hence, provide much information for determining these properties. Importantly, due to the correlation between $R_{1.4}$ and slope parameter $L$, this nuclear empirical parameter is also well constrained. By using the universal relation between neutron skin thickness $R_{\rm skin}^{208}$ and slope $L$ \citep{2014EPJA...50...27V}, we have transferred the astrophysical constraints on $L$ to that of $R_{\rm skin}^{208}$. The 68.3\% credible level uncertainty is $R_{\rm skin}^{208}=0.204^{+0.030}_{-0.026}\,{\rm fm}$, which is consistent with the results of, e.g., Refs.~\citep{2021arXiv210502886B, 2021arXiv210210074E, 2021Univ....7..182L}. Obviously, this result is still much smaller than the PREX-II measurement but slightly mitigates the tension between measurements of PREX-II and other nuclear experiments. It is also noticeable from the fitting results (see Fig.~\ref{fig:mrfit}) that the reconstructed radii of GW170817 favor the second peak of the distribution of previous results \citep{2018PhRvL.121p1101A}, and this also relaxes the tension between PREX-II measurement and previous small tidal deformability measurements. By comparing our constraints on maximum mass of NS with previous $M_{\rm TOV}$ determinations in , e.g., Refs.~\citep{2020ApJ...904..119F, 2020PhRvD.102f3006S}, we find that current observational and theoretical knowledge of the NS population and its EOS, favors a binary black hole merger scenario for GW190814.

However, due to the relatively large uncertainties of both PREX-II and NICER measurements, as well as the loose $M_{\rm TOV}$ constraints, the conclusion made in this work still needs to be checked with future observations. For example, the determinations of stellar radii by NICER for NSs with known masses, such as {PSR J0437-4715}, could be made at a $\pm3$\% level \citep{2019ApJ...887L..27G}. Meanwhile, KAGRA will join the LIGO and Virgo detector networks in the fourth observing run, and benefiting from the improved sensitivities and detector numbers, the detected events will be largely increased, in which catching another BNS merger is promising. Hence, with our hybrid parameterization method that is compatible with nuclear empirical expansion and able to resemble various theoretical EOS models, the existence of the tensions discussed above will be further probed, and then we will be able to shed valuable light on the dense matter EOS as well as the nuclear symmetry energy.

\begin{acknowledgments}
We thank the anonymous referee for the helpful suggestions. This work was supported in part by NSFC under Grants No. 11921003, No. 11933010, and No. 12073080, as well as the Chinese Academy of Sciences via the Strategic Priority Research Program (Grant No. XDB23040000) and the Key Research Program of Frontier Sciences (No. QYZDJ-SSW-SYS024).
\end{acknowledgments}

\appendix
\begin{figure*}
    \centering
    \includegraphics[width=0.75\textwidth]{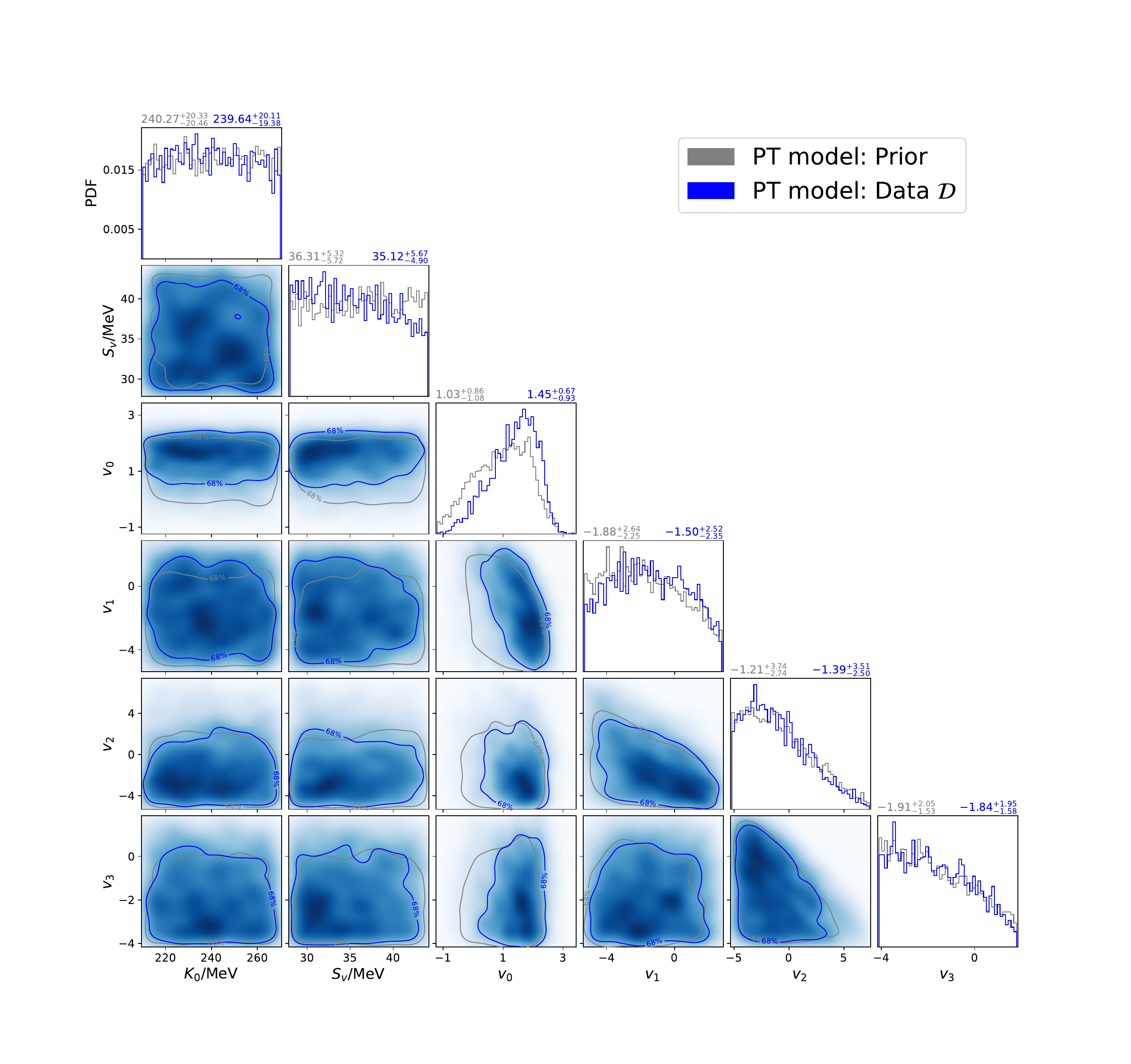}\vspace{-8mm}
    \includegraphics[width=0.75\textwidth]{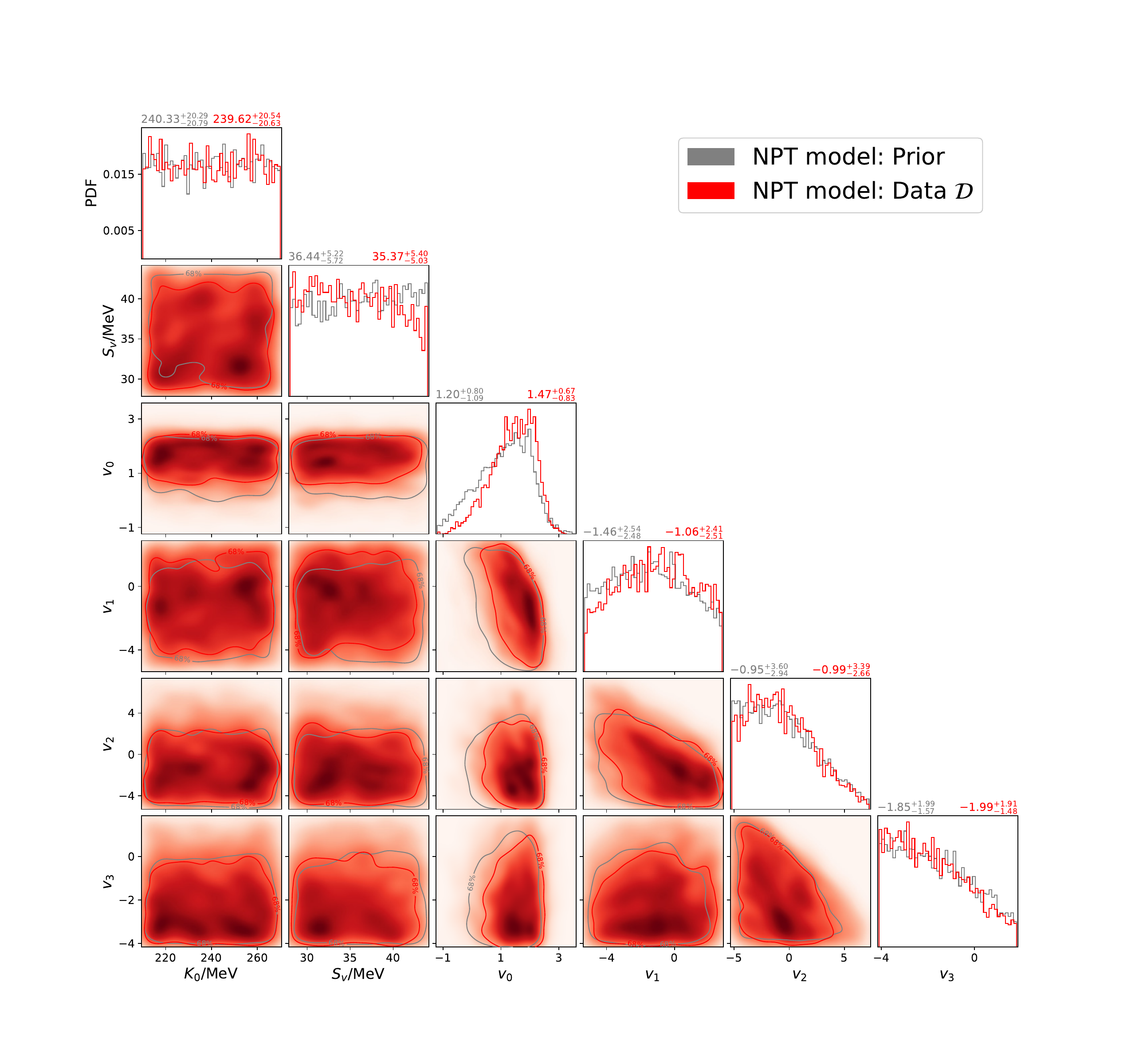}\vspace{-8mm}
    \caption{Corner plots of posteriors for EOS parameters $\{K_0, S_v, v_0, v_1, v_2, v_3\}$.}
    \label{fig:vk}
    \hfill
\end{figure*}

\bibliographystyle{apsrev4-1}
\bibliography{bibtex}

\end{document}